# Topologically protected acoustic helical edge states and interface states in strongly coupled metamaterial ring lattices

Xue-Feng Zhu[*†1,4], Yu-Gui Peng[†1,4], Xiang-Yuan Yu[2,3], Han Jia[*2], Ming Bao[2], Ya-Xi Shen[1], and De-Gang Zhao[*1]

[1]Department of Physics, Huazhong University of Science and Technology, Wuhan, 430074, China

[2]Key Laboratory of Noise and Vibration Research, Institute of Acoustics, Chinese Academy of Sciences, Beijing 100190, China

[3]School of Automation, Wuhan University of Technology, Wuhan 430063, China

[4]Innovation Institute, Huazhong University of Science and Technology, Wuhan, 430074, China

[†]These authors contributed equally to this work.

[*]e-mails: xfzhu@hust.edu.cn, hjia@mail.ioa.ac.cn, and dgzhao@hust.edu.cn.

Time reversal (*T*) invariant topological insulator is widely recognized as one of the fundamental discoveries in condensed matter physics, for which the most fascinating hallmark is perhaps a spin based topological protection, the total cancellation of scattering of conduction electrons with certain spins on matter surface[1-5]. Recently, it has created a paradigm shift for topological insulators, from electronics to photonics[6-12], phononics[13,14] as well as mechanics[15-17], bringing about not only involved new physics but also potential applications in robust wave transport. Despite the growing interests in realizing topologically protected acoustic wave transport[18], *T*-invariant acoustic topological insulator has not yet been achieved. Here, we report the first demonstration of acoustic topological insulator: a strongly coupled metamaterial ring lattice that supports one-way propagation of helical edge states under *T*-symmetry, backscattering immune to boundary abrupt variations. The very unique thing is the formation of spin-filtered interface states due to lattice dislocations. The mechanism underlying the formation of topologically protected edge states and interface states is applicable in various other wave systems or higher dimensions.

The study of topological properties or invariants in periodic physical systems has become one of the most active fields in many branches of physics[1-20]. As basic elements of topological band theory, these topological invariants are very interesting because they imply the presence of nontrivial bulk topologies, giving rise to the presence of gapless edge or surface states featured with the absence of backscattering. Probably the most well-known example should be the integer quantum Hall effect discovered by Klaus von Klitzing *et al.* under strong magnetic fields and low temperatures[21], where the quantized Hall conductivity is insensitive to local perturbations, and the quantization (TKNN integer) was afterwards discovered to have an expression of the first Chern number of the bands below the chemical potential[22]. Another seminal work is the Kane and Mele model, dealing with the time reversal ($T$) invariant systems of strong spin-orbit couplings[23]. In that work, Kane *et al.* introduce a distinctive $Z_2$ index to describe the $T$-invariant quantum spin Hall phase that has a spin-dependent topological protection and robustness against non-magnetic disorder and interactions under $T$-symmetry. The new topological phase in presence of $T$-symmetry was later called topological insulators, now widely recognized as a fundamental discovery in condensed matter physics[1-5]. In the past decade, the exploration for topological invariants is substantively followed up in different subfields of physics, making a paradigm shift for topological states, from electronics to photonics[6-12], phononics[13,14] as well as mechanics[15-17]. In acoustics, unidirectional edge channels have been recently proposed for scalar acoustic waves propagating in a fluid circulation array[18], mimicking the integer quantum Hall effect by breaking *T*-symmetry via the Doppler effect. However, the acoustic equivalent of topological insulators, as a missing part corresponding to the *T*-invariant quantum spin Hall phase, has still not yet been achieved. As revealed in the Kane and Mele model[23], the main challenges for the realization of spin-dependent scatter-free acoustic behaviors root in creating decoupled spins under *T*-symmetry and strong spin-orbit couplings.

In this report, we have theoretically proposed and experimentally demonstrated the acoustic topological insulator. The analogy between our proposed acoustic topological insulator and quantum spin Hall effect is schematically presented in Fig. 1. Here we define a pseudo spin for acoustic waves propagating in the strongly coupled ring lattice based on wave circulation direction at the lattice points[7,8,11], *viz.* spin up (↑) ↔ clockwise and spin down (↓) ↔ anti-clockwise. As shown in previous works[24], the fundamental distinction in *T*-symmetry between phonons and electrons is that phonons with different pseudo spins are easily converted into each other through time-reversed channels, while electrons with different spins are completely decoupled under *T*-symmetry protection. Therefore, phonons can hardly have the same 2D or 3D topological insulators for electrons. In this work, we realize decoupled spins under *T*-symmetry for acoustic waves and strong spin-orbit coupling through nearly unitary coupling strength ($|\kappa|^2 > 0.9$) in the ring lattice. As a result, acoustic waves with a pseudo spin in one ring will tunnel into the neighboring link ring with the pseudo spin flipped, and then go to the next ring with the pseudo spin restored, rendering a "zigzag" route having two sets of topologically protected connections where no waves make a trip. By purposely designing the lattice configuration, we can utilize those topologically protected connections and unitary tunneling at each coupling point to construct decoupled helical edge states as well as interface states with backscattering immune to boundary abrupt variations or lattice distortion.

As shown in Fig. 1, we can see that the direction of momentum and pseudo spin of acoustic waves are locked at the boundaries of the acoustic topological insulator. For example, the upper edge supports forward propagating waves with spin up (red arrows) and backward propagating waves with spin down (blue arrows) and conversely for the lower edge. In the ring lattice, we define the numbers of helical edge states propagating in clockwise and anti-clockwise to be $N_\uparrow$ and $N_\downarrow$, respectively. According to Laughlin argument[25], the Chern number $C$ is equal to the net number $N_\uparrow - N_\downarrow$ of helical edge states. For $T$-symmetric systems, the equivalent number of different helical edge states ($N_\uparrow = N_\downarrow$) ensures a zero Chern number $C=0$. However, when the two spin components are decoupled, we can define spin Chern numbers for spin up ($C_\uparrow$) and spin down ($C_\downarrow$) components, respectively. In our case, $C_\uparrow = 1$ ($N_\uparrow = 1$, $N_\downarrow = 0$) and $C_\downarrow = -1$ ($N_\uparrow = 0$, $N_\downarrow = 1$) at the tunneling frequencies. The standard $Z_2$ index $v$ is then given by[5,23]

$$v = \frac{C_\uparrow - C_\downarrow}{2} \mod 2 = 1, \tag{1}$$

which is a nontrivial bulk topological invariant. For each spin component, the clockwise (or anti-clockwise) propagation of helical edge states indicates the existence of an effective magnetic field in the normal direction[7,8,11,24]. When the helical edge states travel through a close loop, the accumulated phase change is proportional to the total field flux threading the loop.

Acoustic topological insulator, like the electronic counterpart, is a strongly coupled physical system. In our work, we first focus on studying the scattering properties of the coupling connection between two U shape waveguides displayed in Fig. 2a, which is also the fundamental coupling element of the acoustic topological insulator shown in Fig. 1. We employ extremely anisotropic metamaterials to support acoustic guiding wave propagation with a large wave vector, because natural materials always possess lower acoustic refractive indices than that of air[26,27]. The metamaterials can be simply constructed by alternating subwavelength air-metal layers (metal: Aluminum alloys), where $w$ and $t$ are the width and thickness of metal plates, and $p$ is the period of structural change, shown by the inset of Fig. 2a. The effective refractive index $n_{eff}$ of the waveguide metamaterials is[26,27]

$$n_{eff} = \sqrt{F^2 \tan^2\left(\frac{w\omega}{2c_{air}}\right) + n_{air}}, \quad (2)$$

which is always larger than the refractive index of air $n_{air}$ ($n_{air}=1$). Here, $F(=1-t/p)$ is the filling ratio of air layers, $\omega$ is the round frequency of acoustic waves, and $c_{air}$ is the speed of sound in air at room temperature ($c_{air}$=343.21 m/s). For the fabricated metamaterials sample, we have $t$=0.004 m, $p$=0.01 m, $F$=0.6, and $w$=0.04 m. From Eq. (2), it requires $\omega<\pi c_{air}/w$ to suppress high-order waveguide modes. When we put two U shape metamaterial waveguides in close proximity (spacing: 0.02 m) as shown in Fig. 2a, the acoustic waves circulating in one direction in a U shape waveguide will be phase-matched to one of the two degenerate counter-propagating modes in the adjacent waveguide, and unitary acoustic tunneling occurs. The scattering process between two coupled U shape waveguides can be described by[28]

$$\begin{bmatrix} p'_3 \\ p'_4 \end{bmatrix} = \begin{bmatrix} t & \kappa \\ -\kappa^* & t^* \end{bmatrix} \begin{bmatrix} p_1 \\ p_2 \end{bmatrix}, \quad \begin{bmatrix} p'_1 \\ p'_2 \end{bmatrix} = \begin{bmatrix} t & \kappa \\ -\kappa^* & t^* \end{bmatrix} \begin{bmatrix} p_3 \\ p_4 \end{bmatrix}, \quad (3)$$

where $t$ and $\kappa$ are the dimensionless transmission and coupling coefficients, $p_i$ refers to the pressure field component. According to conservation of energy, the coupling matrix should be unimodular so that $|t|^2+|\kappa|^2=1$[28]. From Eq. (3), if the acoustic waves are only input at Port 1 (e.g., $p_1=1, p_2=p_3=p_4=0$), we will obtain $p'_1=p'_2=0$, $p'_3=t$, and $p'_4=-\kappa^*$. The normalized output energies at Ports 2-4 are thus $I_2$=0, $I_3$=$|t|^2$, and $I_4$=$|\kappa|^2$, respectively. As $|\kappa|\to1$, we will further have $I_3\to0$ and $I_4\to1$. In this case, the coupled waveguide mode can actually be regarded as a conventional waveguide mode. To search for the frequency range where U shape waveguides are strongly coupled, we have calculated the transmitted sound energy at Ports 2-4 from 2700 Hz to 3500 Hz (Fig. 2b), where the acoustic guiding waves are input at Port 1. In Fig. 2b, we note that the sound energy at Port 2 is nonzero ($I_2\neq0$), which is attributed to the little reflections at the output facets of Ports 3 and 4 with mismatched acoustic impedances. The sound energy spectrum at Port 4 shows that the coupling strength is close to unitary ($|\kappa|^2>0.9$) from 2940 Hz to 3090 Hz, and the maximum value is ~0.94 at 3020 Hz. To verify the acoustic tunneling effect, we simulate the sound energy distribution in the coupled metamaterial waveguides at 3020 Hz by using finite element solver (COMSOL Multiphysics), plotted in Fig. 2c. The simulation clearly shows that the input sound from Port 1 is completely tunneling through the coupling connection and output at Port 4. Since now the coupled waveguide mode behaves like a conventional waveguide mode, the coupling connection can principally be anywhere on the curved parts of the U shape waveguides (see Supplementary Information). The nearly unitary coupling strength will consequently lead to a high contrast between the sound energy at Port 4 and the one at Port 2 (or Port 3). To make a demonstration, we carefully measured the sound energy at Ports 2-4 in the frequency range from 2940 Hz to 3090 Hz and calculated the energy contrasts of $I_4/I_2$ and $I_4/I_3$, respectively. In the experiment, the fabricated metamaterial waveguides are sealed in rigid rectangular waveguides to prevent unwanted radiation losses during the propagation and isolate environmental noises (Fig. 2a). The results in Fig. 2d show that the measured energy contrasts can reach up to ~20 at the frequencies of our interest, in a qualitative agreement with the simulations.

Next, we add a metamaterial ring resonator between the two U shape waveguides to demonstrate the essence of acoustic topological insulator, *viz*. the formation of topologically protected connections. Figure 3a shows the fabricated U-Ring-U shape metamaterial waveguide sample, which is a four port system with two coupling connections. For comparison, we perform the full wave simulation of sound energy distribution in the structure at 3020 Hz (Fig. 3b). It is evident that the guiding wave input at Port 1 is "zigzagging" through the metamaterial ring via unitary tunneling at the two coupling connections and output at Port 2. In this case, as aforementioned, the metamaterial ring no longer acts as a resonator and the coupled waveguide mode is essentially a conventional waveguide mode (see Supplementary Information). Here, the very unique thing is that the guiding wave is passing through the metamaterial ring by making only a half round trip, thus forming a dark section (marked by the dashed circle in Fig. 3b) in the ring disabled in conducting energy through additional coupling channels. We call the coupling connections in the dark section to be topologically protected, which, as will be shown later, plays a dominant role in realizing acoustic topological insulator. For the demonstration, we measured sound energy distribution in the metamaterial ring from 2940 Hz to 3090 Hz, as shown in Fig. 3c (see Supplementary Information). We observe that the sound energy is mainly distributed along the lower half circle, leaving the upper half circle almost dark at the frequencies of our interest. For a more intuitive presentation, we replot the data on the vertical dashed line (Fig. 3c) at 3020 Hz in Fig. 3d and compare it with the simulation result in Fig. 3b. The sound energy distributions in Figs. 3b and 3d are in fair agreement, verifying the formation of topologically protected connections in the upper half circle, while the manifestation of pressure field attenuation in the propagation may be resulted from the loss of thermal damping in the subwavelength slits of the metamaterials and radiation loss from sample imperfection. We can also obtain the energy contrast at 90° and 270° positions in the metamaterial ring from 2940 Hz to 3090 Hz in Fig. 3e, using the data on horizontal dashed lines (Fig. 3c). The results show that the measured energy contrasts are obvious from 2940 Hz to 3090 Hz and maximally take the values of up to ~60 at the frequencies round 3016 Hz. Basically, we can make an extension in designing a coupled metamaterial ring lattice, where the coupling connections between the interior rings and the boundary rings are all topologically protected. Therefore, it is expected that the coupled ring lattice will support the reflectionless "zigzag" waves propagating at the boundaries and exhibit similar intriguing properties of electronic topological insulators as described previously in Fig. 1.

In the last, we numerically display the representative properties of acoustic topological insulators with different lattice configurations. The simulations are operated in lossless systems. We point out those properties are observable even in the presence of decay processes as revealed in Fig. 3. The first demonstration is the one-way pseudo spin dependent transport of edge states for a lattice of 2×3 rings unit-cells inserted with an elliptic rigid defect. The U shape waveguide is imposed to selectively excite (or extract) spin up and spin down acoustic helical edge states. When we launch the spin up helical edge state (Fig. 4a), the guiding wave is propagating in clockwise and perfectly transmitted by the six sharp corners of the lattice. Due to the nearly unitary coupling at each connections, the reflection into the backward propagating spin down component is little and the output sound energy is closely equal to the input. When we launch the spin down helical edge state, the guiding wave is propagating in anti-clockwise along a different route with only one sharp corner, shown in Fig. 4b. We also explore the case of interfacial defects, where two lattices of 1×3 rings unit-cells are horizontally dislocated by half a lattice constant. It is found that the spin up helical state launched on the left could flow along the interface with little reflections (Fig. 4c). In this case, we note that the two sets of coupling channels between the interfacial rings and the interior rings of two lattices are all disabled (or topologically protected). However, if we launch the spin down helical state on the left (Fig. 4d), the guiding wave turns into a bulk state with the sound energy distribution exactly complementary to that of the interface state shown in Fig. 4c.

In this work, we have demonstrated the first acoustic topological insulator in analogy with the quantum spin Hall effect, which permits unidirectionally propagating helical edge states and spin-filtered helical interface states. Our work provides a fertile ground for robust sound transports and fundamental explorations of topological acoustics. Of interest will be the extension of our work into nonreciprocal acoustics regime by integrating time-varying, which shed lights on the developments of chiral acoustic metamaterials and Chern acoustic topological insulator with sound isolation and black-hole harvesting properties.

**ACKNOWLEDGEMENTS**

This work was supported by the National Natural Science Foundation of China under No.11404125, No.11304351, No.11304105, and the "Strategic Priority Research Program" of the Chinese Academy of Sciences (Grant No. XDA06020201). X. F. Z. acknowledges the financial support from the Bird Nest Plan of HUST.

**AUTHOR CONTRIBUTIONS**

X. F. Z. and Y. G. P carried out the numerical simulation and theoretical analysis. H. J., Y. G. P., and X. Y. Y. fabricated the sample. H. J., Y. G. P., and X. Y. Y. performed the measurement and the data processing. X. F. Z. conceived the idea. M. B. and D. G. Z. helped in the discussion. X. F. Z. and H. J. designed the experiment and supervised the study. All authors contributed to the preparation of the manuscript.

**ADDITIONAL INFORMATION**

Competing financial interests: The authors declare no competing financial interests.

**FIGURES AND CAPTIONS**

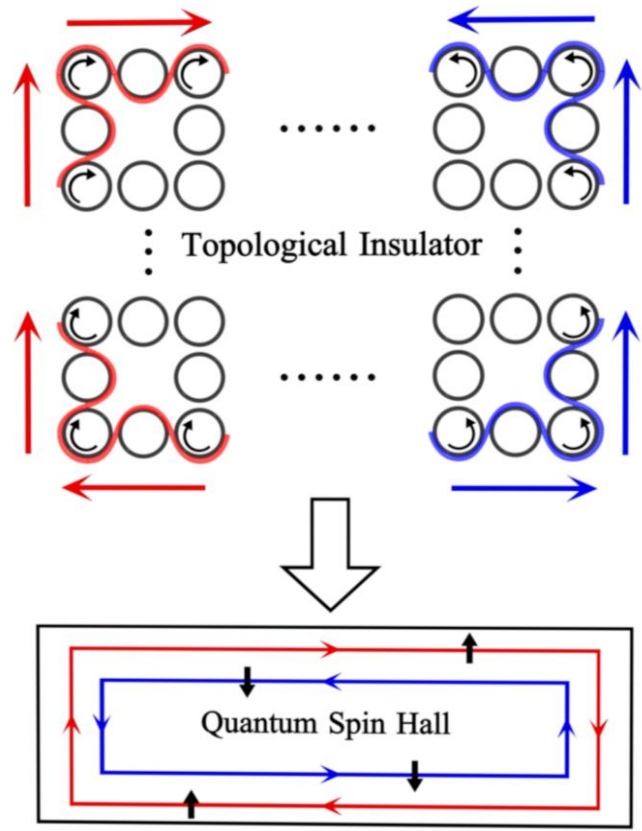

Figure 1. **Analogy between acoustic topological insulator (a strongly coupled ring lattice) and quantum spin Hall effect.** The spinful 2D system has 4 basic degrees of freedom. The edges support a clockwise propagating state with spin up and an anti-clockwise propagating state with spin down. The edge states with different spins are regarded as decoupled due to nearly unitary coupling strength between neighboring rings, closing the time-reversed channels for back-reflections.

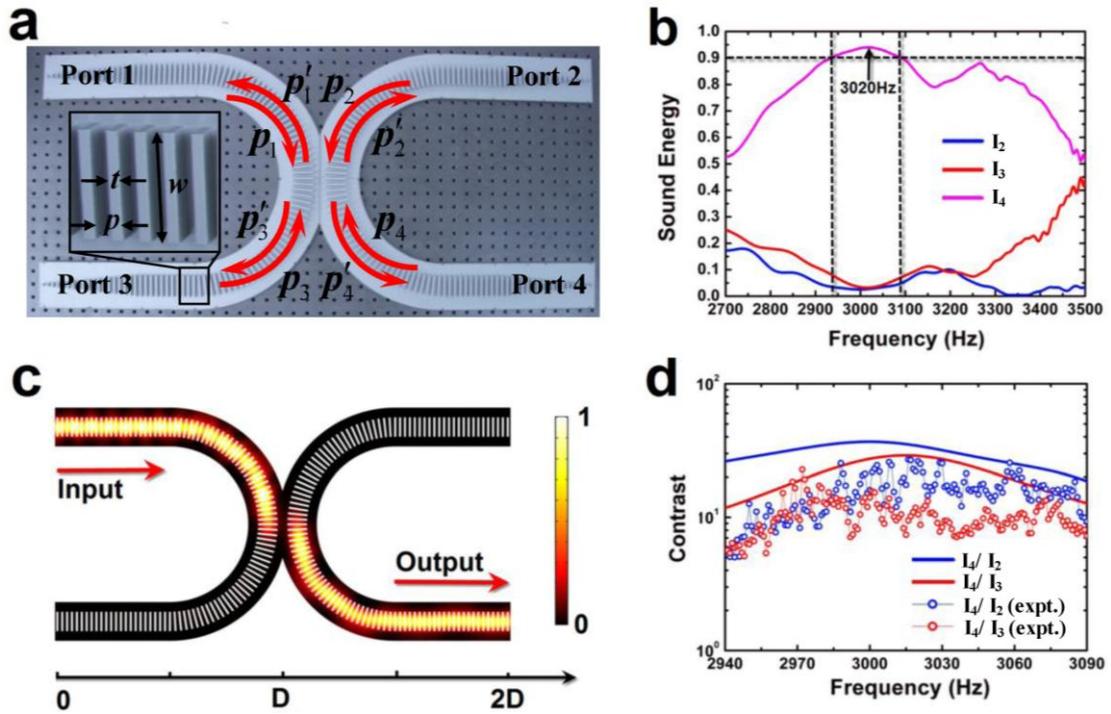

Figure 2. **Demonstration of the nearly unitary coupling strength between two adjacent U shape waveguides. a**, Photos of two coupled U shape metamaterial waveguides. The structure can be regarded as a four-port system, where the forward and backward propagating pressure field components are labeled. The metamaterials comprise an array of metal plates spaced by air gaps. **b**, The transmitted sound energy spectra at Port 2 ($I_2$), Port 3 ($I_3$), and Port 4 ($I_4$), respectively. It can be seen that the coupling strength $|\kappa|^2$ is over 0.9 in the frequency range from 2940 Hz to 3090 Hz and reaches maximum (~0.94) at 3020 Hz. **c**, Full-wave simulation of sound energy distribution in the coupled metamaterial waveguides at 3020 Hz, where the guiding wave is input at Port 1. Due to the nearly unitary coupling strength, the guiding wave is tunneling into the adjacent waveguide and output at Port 4. In **c**, the scale mark $D$=0.46 m. **d**, The calculated (solid lines) and measured (circled lines) energy

contrasts of $I_4/I_2$ and $I_4/I_3$, showing most sound energy are output at Port 4 in the frequency range from 2940 Hz to 3090 Hz.

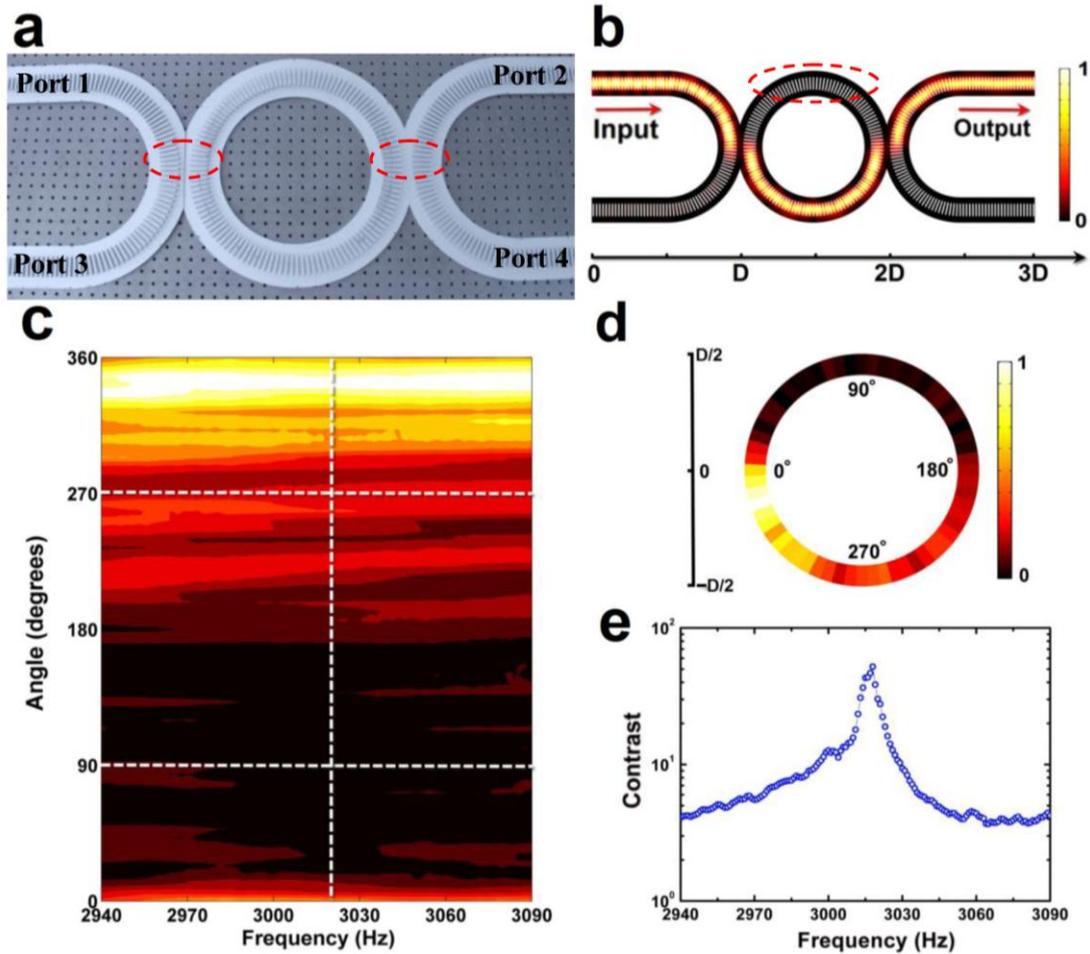

Figure 3. **Demonstration of the formation of topologically protected connections.** **a**, Photos of the U-Ring-U shape metamaterial waveguides. The structure is a four-port system with two coupling connections marked by the dashed circles. **b**, Full-wave simulation of sound energy distribution in the structure at 3020 Hz, where the guiding wave is input at Port 1. Due to the nearly unitary coupling strength, the guiding wave will "zigzag" through the ring without making a complete round-trip and be output at Port 2. **c**, Measured sound energy distribution in the metamaterial

ring from 2940 Hz to 3090 Hz. Here the vertical dashed line shows the measured sound energy distribution in the metamaterial ring at 3020 Hz (**d**). From the horizontal dashed lines, we can obtain the energy contrast at 90° and 270° positions in the metamaterial ring from 2940 Hz to 3090 Hz (**e**).

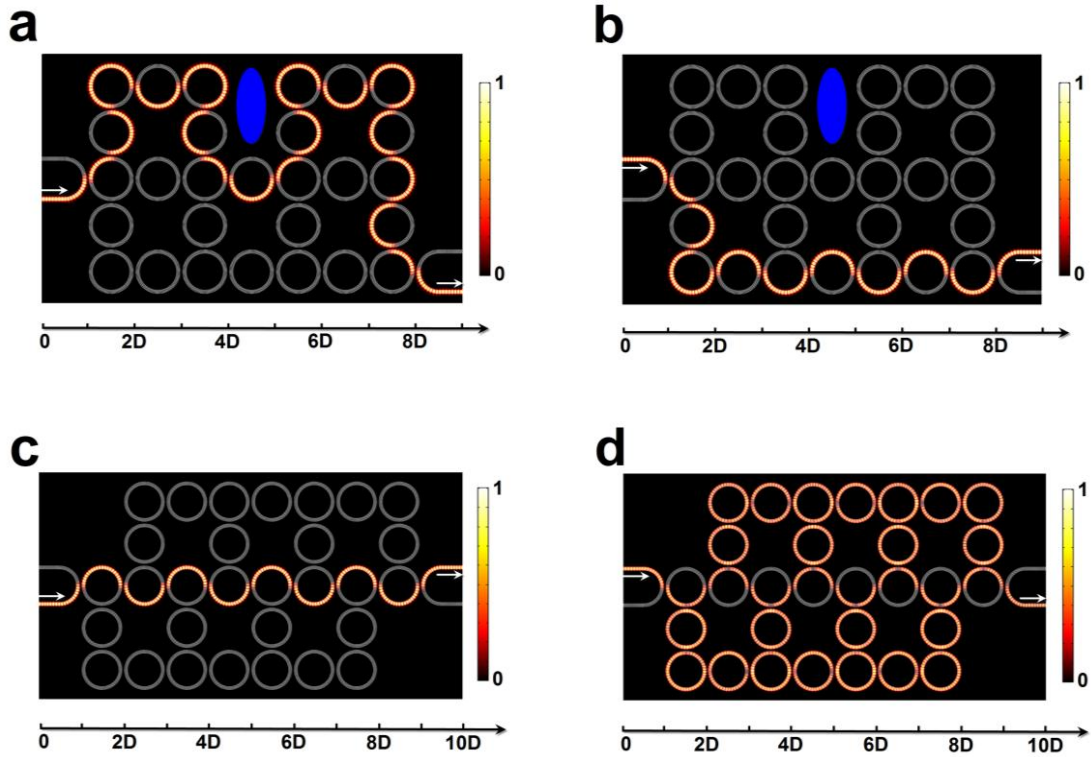

Figure 4. **Properties of the acoustic topological insulator.** Selective excitation of spin-up (**a**) and spin down (**b**) acoustic one-way edge states. We can clearly observe the robustness of the edge states against the sharp bending of the boundaries. The elliptical cylinder in **a** and **b** represents a rigid defect. After introducing a lattice dislocation by half a lattice constant, we obtain a novel acoustic one-way interface state in **c**, propagating from left to right for the spin-up component, and conversely for the spin-down component. If we excite the wrong spin component at the input port,

we will obtain a bulk state instead of an interface state as shown in **d**.